\documentclass[nohyperref]{article}

\usepackage{microtype}
\usepackage{graphicx}
\usepackage{subfigure}
\usepackage{booktabs} 
\usepackage{apxproof}
\usepackage{multicol}

\newlength{\RoundedBoxWidth}
\newsavebox{\GrayRoundedBox}
   {\setlength{\RoundedBoxWidth}{\dimexpr#1}
    \begin{lrbox}{\GrayRoundedBox}
       \begin{minipage}{\RoundedBoxWidth}}%
   {   \end{minipage}
    \end{lrbox}
    \begin{center}
    \begin{tikzpicture}%
       \draw node[draw=black,fill=black!10,rounded corners,%
             inner sep=2ex,text width=\RoundedBoxWidth]%
             {\usebox{\GrayRoundedBox}};
    \end{tikzpicture}
    \end{center}}
    
\usepackage{hyperref}

\newcommand{\bb}[1]{\textbf{\textsf{#1}}}

\usepackage[accepted]{icml2022}

\usepackage{amsmath}
\usepackage{amssymb}
\usepackage{mathtools}
\usepackage{amsthm}

\usepackage[capitalize,noabbrev]{cleveref}

\theoremstyle{plain}
\newtheorem{theorem}{Theorem}[section]

\theoremstyle{definition}

\newtheorem{assumption}[theorem]{Assumption}
\theoremstyle{remark}
\newtheorem{remark}[theorem]{Remark}

\usepackage[textsize=tiny]{todonotes}

\def \bepsilon{\mbox{\boldmath$\epsilon$}}
\def \by{\mbox{\boldmath$y$}}
\def \bx{\mbox{\boldmath$x$}}

\def\R{{\mathbb R}}
\def\E{{\mathbb E}}

\def \btheta{\boldsymbol{\theta}}

\def \bx{\boldsymbol{x}}

\def \bD{\boldsymbol{D}}

\def \bA{\boldsymbol{A}}
\def \bI{\boldsymbol{I}}

\def \bz{\boldsymbol{z}}
\def \bPsi{\boldsymbol{\Psi}}
\def \bPhi{\boldsymbol{\Phi}}

\def \bLambda{\boldsymbol{\Lambda}}
\def \bpsi{\boldsymbol{\psi}}
\def \btheta{\boldsymbol{\theta}}
\def \bphi{\boldsymbol{\phi}}

\def \bDelta{\boldsymbol{\Delta}}

\definecolor{myred}{rgb}{1,0,0.25}

\def \st{s}

\icmltitlerunning{Filtered Iterative Denoising}

\begin{document}

\twocolumn[
\icmltitle{Filtered Iterative Denoising for Linear Inverse Problems}



\icmlsetsymbol{equal}{*}

\begin{icmlauthorlist}
\icmlauthor{Danica Fliss}{yyy}
\icmlauthor{Willem Marias}{zzz}
\icmlauthor{Robert D. Nowak}{yyy}
\end{icmlauthorlist}

\icmlaffiliation{yyy}{Department of Electrical and Computer Engineering, University of Wisconsin-Madison}
\icmlaffiliation{zzz}{Space Science and Engineering Center, University of Wisconsin-Madison}

\icmlcorrespondingauthor{D. Fliss}{dfliss@wisc.edu}

\icmlkeywords{Machine Learning, ICML}

\vskip 0.3in
]



\printAffiliationsAndNotice{}  

\begin{abstract}
Iterative denoising algorithms (IDAs) have been tremendously successful in a range of linear inverse problems arising in signal and image processing.  The classic instance of this is the famous Iterative Soft-Thresholding Algorithm (ISTA), based on soft-thresholding of wavelet coefficients.  More modern approaches to IDAs replace soft-thresholding with a black-box denoiser, such as BM3D or a learned deep neural network denoiser. These are often referred to as ``plug-and-play" (PnP) methods because, in principle, an off-the-shelf denoiser can be used for a variety of different inverse problems.  The problem with PnP methods is that they may not provide the best solutions to a specific linear inverse problem; better solutions can often be obtained by a denoiser that is customized to the problem domain.  A problem-specific denoiser, however, requires expensive re-engineering or re-learning which eliminates the simplicity and ease that makes PnP methods attractive in the first place.  This paper proposes a new IDA that allows one to use a general, black-box denoiser more effectively via a simple linear filtering modification to the usual gradient update steps that accounts for the specific linear inverse problem.  The proposed Filtered IDA (FIDA) is mathematically derived from the classical ISTA and wavelet denoising viewpoint.  We show experimentally that FIDA can produce superior results compared to existing IDA methods with BM3D.
\end{abstract}

\section{Introduction}
\label{intro}

Black-box denoisers like BM3D and deep neural network denoisers form the backbone of state-of-the-art methods for solving linear inverse problems in signal and image processing.  Using denoisers for regularization is attractive because one can use an off-the-shelf denoiser in a variety of different inverse problems, sometimes referred to as ``plug-and-play" methods. We argue, however, that the regularization should be adapted to the linear operator of the forward problem.  This would require re-learning a denoiser for each specific linear inverse problem, defeating the simplicity and flexibility of such approaches.  To circumvent this, we propose a novel approach that instead appropriately modifies the data-fitting objective and leads to a filtered gradient update, eliminating the need for learning or adaptation of the denoiser.  We call our new approach a Filtered Iterative Denoising Algorithm (FIDA).

This paper considers the following form of \emph{linear inverse problem}. Let $\by$ denote observations of a signal or image $\bx$ given by 
\begin{eqnarray}
    \by & = & \bA \bx +  \bepsilon
    \label{eq:lininv}
\end{eqnarray}
where $\bA$ is a known linear operator and $\bepsilon = \by -\E[\by]$ may be viewed as a mean zero noise. In other words, we assume that the expected value of $\by$ is a linear transformation of $\bx$. The linear operator $\bA$ can denote the effect of blurring, subsampling, compressed sensing, tomographic projection, or other distortions.  Throughout the paper, we assume that $\bx$ and $\by$ are real-valued vectors and $\bA$ is a real-valued matrix with compatible dimensions (extensions to complex-valued objects are possible). The goal is to recover $\bx$ from the data $\by$. The recovery problem is often ill-posed and regularization methods are used to find a solution that balances the fit to the data and the regularity of the solution (measured in an appropriate sense).

\section{Iterative Denoising Algorithms}

Consider what we will call an \emph{Iterative Denoising Algorithm} (IDA), outlined in Algorithm~\ref{IDA} below.  Let $L$ denote a loss function measuring the quality of a solution $\bx$.  This paper will focus on the squared error loss $L(\bx) = \frac{1}{2}\|\by -\bA \bx\|_2^2$, but extensions to other losses may be possible.  IDA iterates between a gradient descent step on the loss followed by a denoising step, denoted by \bb{\small denoise} that effectively regularizes each iterate. The denoiser takes the  gradient descent iterate and a denoising parameter $\lambda_\gamma$ as inputs and outputs a denoised version of the iterate.  The parameter $\lambda_\gamma$ may depend on the stepsize $\gamma$ and other problem parameters such as the noise level.


\begin{figure}[h]
    \centering
    \includegraphics[width=8cm]{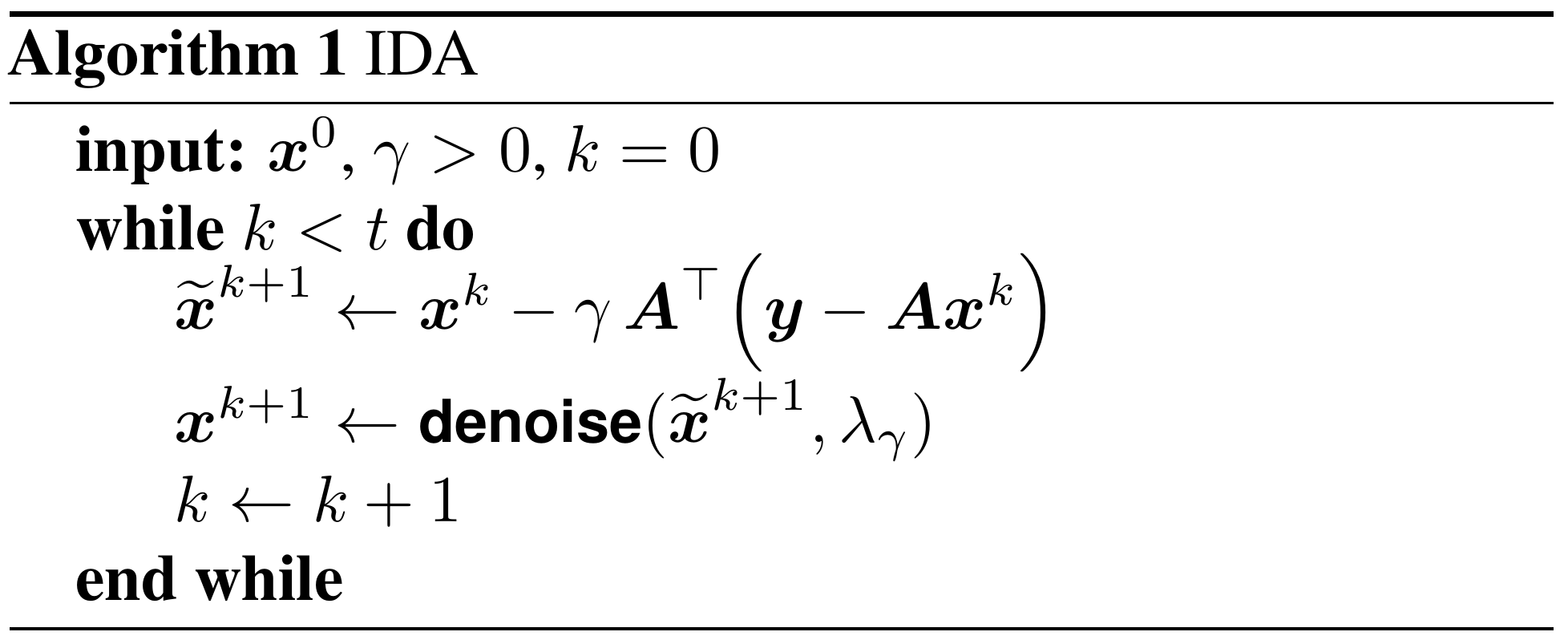} 
    \label{IDA}
\end{figure}

The genesis of such algorithms is traced back to iterative soft-thresholding algorithms \cite{nowak2001fast,figueiredo2003algorithm,daubechies2004iterative,figueiredo2007gradient,wright2009sparse,beck2009fast}. The soft-thresholding operation is the proximal operator for the $\ell^1$ norm regularizer and the threshold level applied at each iteration is proportional to the stepsize $\gamma$.
General IDA methods replace the soft-thresholding denoiser with a off-the-shelf or black-box denoisers such as BM3D \cite{venkatakrishnan2013plug} or deep neural network denoisers \cite{deep-inverse-imaging}.   
IDA and its many variants are often referred to as \emph{plug-and-play} regularization methods because one can simply ``plug-in" any denoiser; see  \cite{kamilov2017plug} for a recent review of such methods.  In this paper, we focus on a basic IDA outlined in (\ref{IDA}), but the main ideas may be extended to related formulations including regularization-by-denoising (RED) \cite{romano2017little}, deep unfolding \cite{chen2018theoretical}, and multiagent consensus equilibrium (MACE) \cite{buzzard2018plug}.

\subsection{The Trouble with Black-Box Denoisers}
\label{trouble}
The trouble with a plug-in denoiser is that it is ignorant of the specific linear operator $\bA$ involved in the inverse problem, which can lead to inappropriate denoising.  To see this, consider a very simple setting where $\bA$ is a square diagonal  matrix with diagonal entries $\delta_i >0$ and assume that $\bx$ is sparse in the canonical basis. Let $\by = \bA\bx + \bepsilon$, where $\bepsilon \sim {\cal N}({\bf 0},\bI)$, a white Gaussian noise vector with variance $1$.  Because $\bx$ is sparse in the canonical basis and $\bA$ is diagonal, it is reasonable to estimate each element of $\bx$ separately.  Note that $y_i \sim {\cal N}(\delta_i x_i, 1)$, so the signal-to-noise ratio  (SNR) in $y_i$ is $\delta_i^2 x_i^2$. The \emph{oracle} denoiser for the $i$-th element of $\bx$ is
\begin{eqnarray*}
    \widehat{x_i}^{O} = \begin{cases}
    y_i, & \text{if $x_i^2>\delta_i^{-2}$}.\\
    0, & \text{otherwise}.
  \end{cases} 
\end{eqnarray*}
In other words, the oracle denoiser "keeps or kills" $y_i$ depending on whether the SNR is greater than $1$.  The key point is that the optimal threshold depends on  $\delta_i$, the $i$-th diagonal element of $\bA$. The analog of the oracle thresholding step would be to apply a hard or soft threshold to $y_i$ itself, and the threshold level should also depend on $\delta_i$.  This observation led \cite{donoho1995nonlinear} to develop the so-called wavelet-vagulette decomposition (WVD), a soft-thresholding algorithm for linear inverse problems that uses varying threshold levels that account for the SNRs induced by the linear operator $\bA$.  {\bf For optimal performance, the denoising step must account for the specific operator involved in the linear inverse problem.}  This tells us that, in general, IDA methods should also adjust the denoiser to the specific $\bA$. The notion that the regularizer or prior may depend on the observation model, while perhaps not widely appreciated, has also been discussed in the Bayesian setting  \cite{gelman2017prior}.

\begin{figure}[h]
    \centering
    \includegraphics[width=8cm]{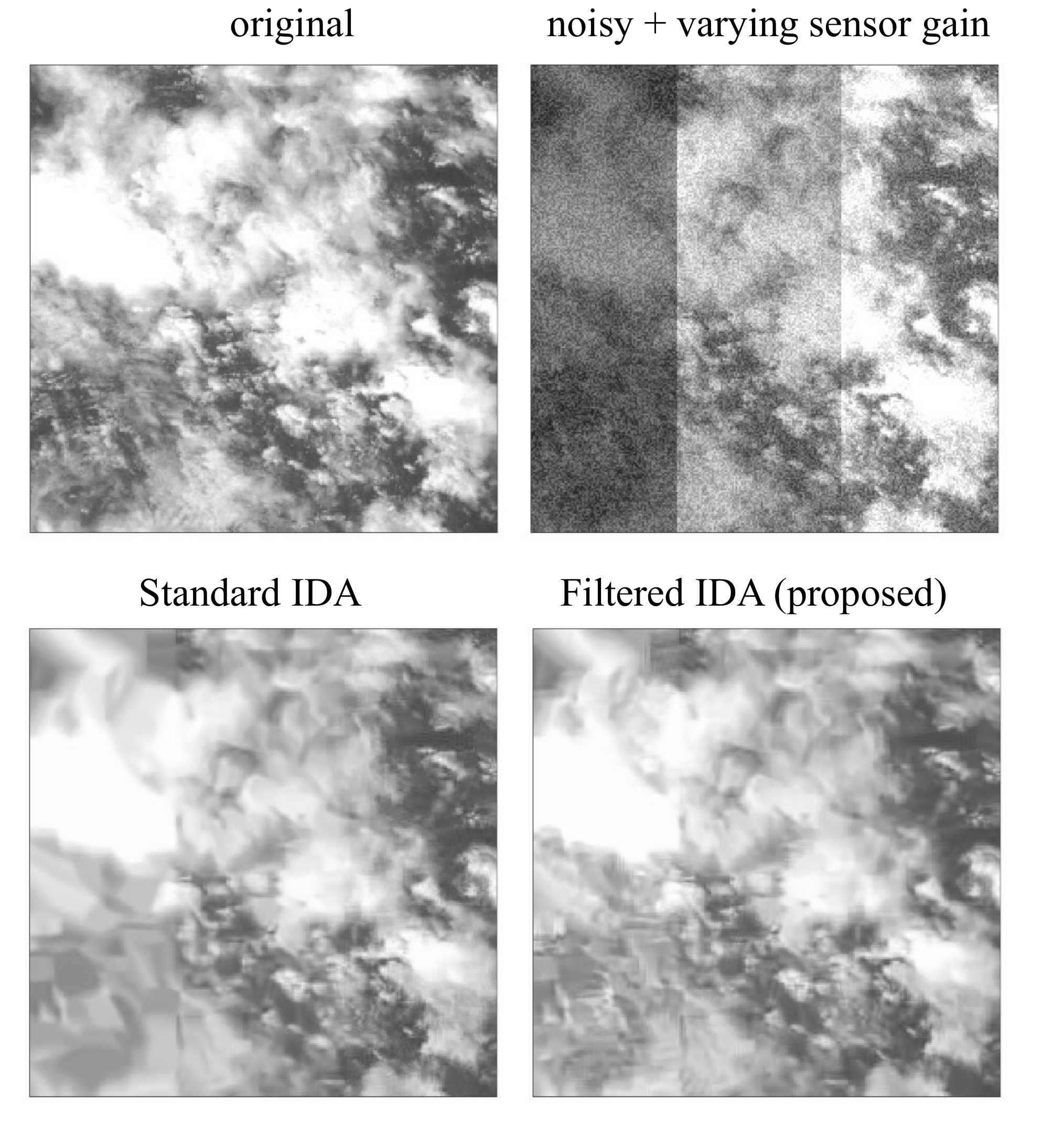} 
    \caption{Satellite image restoration example.  Notice that the new Filtered IDA method does a much better job of denoising and preserving structure, particularly visible in lower left of images, than the standard IDA method.}
    \label{fig:satellite}
\end{figure}

To illustrate the problem, consider the following satellite imaging problem that motivated our investigation. The Visible Infrared Imaging Radiometer Suite (VIIRS) day night band (DNB) is used for cloud type identification via spatial textures.
The utility of DNB observations, however, is highly dependent on the signal to noise ratio (SNR) of the observations which is directly proportional to the lunar luminosity and zenith angle.  Low levels lunar light results in noisy images.  Complicating matters further, light sensors have varying (known) gain factors, which results in the striping artifact shown in Figure~\ref{fig:satellite}. 
The figure also shows results for the standard Iterative Denoising Algorithm (IDA) and the proposed Filtered IDA, using BM3D as the denoiser in both cases.  The BM3D denoising parameter was adjusted separately in both cases to obtain the best results (for fair comparison). 
The filtered IDA method proposed in this paper produces significantly better results, especially noticeable in the the lower left of the images in Figure~\ref{fig:satellite}. 


\section{Filtered IDA}
  The goal is to develop a new IDA strategy that avoids the trouble with black-box denoisers (i.e., avoids the need to adjust the denoiser to the specific linear operator $\bA$). Many black-box denoisers like total variation, BM3D, and deep neural denoisers operate similarly to a wavelet soft-thresholding operation. 
 Total variation is the $\ell^1$ norm of the gradient.  In fact, the connection is even deeper since the $k$th largest Haar wavelet coefficient of an image is upper bounded by the total variation of the image divided by $k$; see Proposition~8 in \cite{needell2013stable}. Block-matching approaches like BM3D \cite{dabov2007image} employ transforms and hard-thresholding (rather than soft-thresholding) operations.
Similarities between iterative soft-thresholding algorithms and multilayer neural network denoisers were explored by \cite{gregor2010learning,xin2016maximal} and others.  These connections support the idea of deriving a new IDA strategy based on $\ell^1$ regularization for sparse models.  The key idea in our new approach will be to modify the gradient descent step rather than the denoiser to account for the specific $\bA$ involved in the problem.

\subsection{Insights from $\ell^1$ Regularization}
 Suppose that $\bPsi$ is an orthogonal matrix constituting a sparsifying orthonormal basis for the signal $\bx$. That is, the coefficient vector $\btheta = \bPsi^\top\bx$ is a sparse or approximately sparse vector.  Iterative soft-thresholding algorithms solve the optimization
$$\min_{\bx} \|\by -\bA\bx\|_2^2 + \lambda \, \|\bPsi^\top\bx\|_1 \ , $$
where $\lambda$ is a regularization parameter that determines the thresholding level.

Our analysis is based on the following key assumption.
\begin{assumption}\label{A1}
The linear operator $\bA$ and the orthogonal matrix $\bPsi$ satisfy the following property. 
$$\bA^\top\bPhi = \bPsi\bDelta $$
where $\bDelta \succ 0$ is a diagonal matrix and $\bPhi$ is a nearly orthogonal matrix satisfying $\bPhi^\top\bPhi \approx \bI$, where $\bI$ is the identity matrix.
\end{assumption}

\begin{remark}
Let $\bpsi_i$ denote the $i$-th column (basis vector) in $\bPsi$.  The assumption states that there exists a complementary vector $\bphi_i$ such that $\bA^\top\bphi_i = \delta_i \bpsi_i$.   
\end{remark}

\begin{remark}
We could assume a more precise quantification of the near-orthogonality of $\bPhi$, but we will only be using the assumption to derive a new iterative denoising algorithm.  We leave quantitative analysis to future work.
\end{remark}

Let $\bpsi_i$  denote the columns (basis vectors) in $\bPsi$. Since $\bPsi$ is an orthogonal matrix, we have the series representation
$$\bx \ = \ \sum_i (\bx^\top\bpsi_i)\,  \bpsi_i \ . $$ Assumption~\ref{A1} allows us to compute each coefficient $\theta_i = \bx^\top\bpsi_i$ from the distorted signal $\bA\bx$ as follows. Let $\delta_i$ denote the $i$-th diagonal element in $\bDelta$ and consider the inner product between $\bA\bx$ and $\bphi_i$, the $i$-th column of matrix $\bPhi$:
$$\bphi_i^\top \bA\bx \ = \ (\bA^\top\bphi_i)^\top\bx \ = \ \delta_i \bpsi_i^\top\bx \ , $$
so $\theta_i = \bpsi_i^\top\bx = \delta_i^{-1} \bphi_i^\top \bA\bx$.  Thus, the same operation may be applied to the noisy observation $\by =\bA\bx+\bepsilon$ (instead of $\bA\bx$) to obtain an unbiased estimator of the coefficient.

\begin{remark}
By Assumption~\ref{A1}, $\bPhi \bPhi^\top\bA = \bPhi \bDelta \bPsi^\top$ and the near-orthogonality of $\bPhi$ then implies $\bA \approx \bPhi \bDelta\bPsi^\top$, a matrix factorization reminiscent of the singular value decomposition.  Since $\bPsi$ is orthogonal, we then have $\bA\bPsi \approx \bPhi\bDelta$ and thus $\delta_i \bphi_i \approx \bA\bpsi_i$ with $\delta_i = \|\bA\bpsi_i\|_2$. 
\end{remark}

\begin{remark}
If $\bA$ is any diagonal matrix and we use the canonical basis $\bPsi=\bI$, then $\bPhi=\bI$.
\end{remark}



Assumption~\ref{A1} is satisfied in a wide variety of situations.  The most well-known is the case where $\bPsi$ is an orthogonal wavelet transform and $\bA$ is a \emph{weakly invertible} linear operator such as an integration operator or Radon transform operator, in which case the factorization is called the \emph{wavelet-vaguelette decomposition} (WVD) \cite{donoho1995nonlinear,abramovich1998wavelet}.  
The WVD method applies a soft-threshold to the unbiased coefficient estimates $\widehat \theta_i =  \delta_i^{-1} \bphi_i^\top \by$, with threshold levels depending on the diagonal entries of $\bDelta$, as derived below.

\subsection{Derivation of Soft-Thresholding IDA} 
First we recall the soft-thresholding function. For $y \in \R$ the solution to the optimization
$$\min_{\theta\in\R} \tfrac{1}{2}(y-\theta)^2 + \lambda |\theta|$$ is given by the soft-thresholding operation 
\begin{eqnarray}\label{soft-threshold}
\widehat \theta &=& \st_\lambda(y) \ := \ \mbox{sign}(y) \max(0,|y|-\lambda) \ . 
\end{eqnarray}

The soft-thresholding denoiser arises from $\ell^1$ regularization, as shown next. Let $\btheta = \bPsi^\top\bx$ and $\bz = \bPhi^\top\by$. Furthermore, let $\bLambda$ be a diagonal matrix with nonzero entries denoted by $\lambda_i$; these will be regularization parameters and determine the thresholding levels. As shown below, it turns out that the thresholding levels should depend on diagonal elements of $\bDelta$.  Consider the weighted $\ell^1$ regularization problem
\begin{eqnarray}
& & \hspace{-.5in} \min_{\bx} \tfrac{1}{2} \|\by -\bA\bx\|_2^2 + \|\bLambda \bPsi^\top\bx\|_1 \nonumber \\
\ & \approxeq & \min_{\btheta} \tfrac{1}{2}  \|\by -\bPhi\bDelta\btheta \|_2^2 + \|\bLambda \btheta\|_1  \nonumber \\
& \approxeq & \min_{\btheta} \tfrac{1}{2} \|\bPhi^\top\by -\bDelta\btheta \|_2^2 + \|\bLambda \btheta\|_1  \nonumber \\
& \equiv & \min_{\btheta} \tfrac{1}{2} \|\bz-\bDelta\btheta \|_2^2 + \|\bLambda \btheta\|_1  \nonumber \\
& \equiv & \min_{\btheta} \tfrac{1}{2} \sum_i (z_i - \delta_i\theta_i)^2 + \lambda_i |\theta_i| \label{opt2} \\
& \equiv & \min_{\btheta} \tfrac{1}{2} \sum_i (\delta_i^{-1} z_i - \theta_i)^2 + \lambda_i \delta_i^{-2}|\theta_i|  \nonumber 
\end{eqnarray}
where Assumption~\ref{A1} is used in the first and second step.  In the final line above, the summation only includes terms when $\delta_i>0$. If $\delta_i=0$, then $\theta_i$ is unrecoverable and our estimate of that coefficient is $0$. The final optimization is separable in each $\theta_i$ and the solution is a soft-thresholding step:
$$\widehat \theta_i \ = \ \mbox{sign}(\delta_i^{-1} z_i) \max(0,|\delta_i^{-1} z_i|-\lambda_i \delta_i^{-2}) \ . $$
The quantity $\lambda_i \delta_i^{-2}$ is the threshold level.  
Recall that $\by = \bA\bx+\bepsilon$, where $\E[\bepsilon]={\bf 0}$.  Let us further make the \emph{white noise} assumption that $\E[\bepsilon\bepsilon^\top] = \sigma^2 {\bf I}$, where $\bI$ is the identity matrix.  Since $\bPhi$ is nearly orthogonal its columns have approximately unit norm. In fact, we may assume the columns each have exactly unit norm by absorbing the normalization factors into $\bDelta$. Therefore, the standard deviation of $\delta_i^{-1} z_i$ is $\delta_i^{-1} \sigma$. Classical statistical arguments \cite{donoho1995nonlinear} dictate a threshold level proportional to the standard deviation of the noise, so take $$\lambda_i = \lambda \, \delta_i$$ for some global $\lambda>0$.  The intuitive explanation for this is obvious: the threshold should be set about the level of the noise standard deviation so that coefficients that are purely noise (which will be many under the sparsity assumption) are set to zero.  This requires threshold levels that are proportional to $\delta_i$, and thus dependent on $\bA$.  In other words, the soft-threshold denoiser must be adjusted based on $\bA$ resulting in the following optimization
\begin{eqnarray}
\min_{\bx} \tfrac{1}{2} \|\by -\bA\bx\|_2^2 + \lambda \, \|\bDelta \bPsi^\top\bx\|_1 
\label{opt1}
\end{eqnarray}
where $\bDelta$ is the diagonal matrix in Assumption~\ref{A1}  and $\lambda>0$ is a regularization parameter.

\section{Filtered Iterative Denoising} 

There is a way around the difficulty of needing to adapt the regularizer/denoiser to the operator $\bA$. Working backwards from (\ref{opt2}) with $\lambda_i = \lambda\delta_i$, we have
\begin{eqnarray*}
&  & \hspace{-.1in} \min_{\btheta}\tfrac{1}{2} \sum_i (z_i - \delta_i\theta_i)^2 + \lambda\delta_i |\theta_i|  \\
&  &  \ \  \equiv \ \min_{\btheta} \tfrac{1}{2} \sum_i (\delta_i^{-1/2} z_i - \delta_i^{1/2} \theta_i)^2 + \lambda |\theta_i|  \\
&  & \ \  \equiv \ \min_{\bx} \tfrac{1}{2} \|\bDelta^{-1/2} \bPhi^\top\by - \bDelta^{1/2} \bPsi^\top\bx\|_2^2 + \lambda \|\bPsi^\top\bx\|_1 
\end{eqnarray*}
which is approximately equivalent to (\ref{opt1}) above (and exactly equivalent if $\bPhi$ is orthogonal, rather than nearly orthogonal). If a particular $\delta_i=0$, then the corresponding diagonal element of $\bDelta^{1/2}$ is also $0$ and we define the corresponding element of $\bDelta^{-1/2}$ to be $0$ as well.. Notice that in the optimization above the denoising regularization term \emph{does not} depend the operator $\bA$.
To minimize this objective using IDA we will compute the gradient of the data-fitting term, which in this case is 
\begin{eqnarray}
\nabla L(\bx) & := & \nabla \tfrac{1}{2} \|\bDelta^{-1/2} \bPhi^\top\by - \bDelta^{1/2} \bPsi^\top\bx\|_2^2 \nonumber \\ 
  & = &  - \bPsi \bDelta^{1/2} \Big(\ \bDelta^{-1/2} \bPhi^\top\by - \bDelta^{1/2} \bPsi^\top\bx\Big) \nonumber \\ & = & \bPsi \bDelta \bPsi^\top\bx -\bPsi  \bPhi^\top\by 
    \label{grad}
\end{eqnarray}

We gain insight into this gradient as follows.
Since 
 $\bA \approx \bPhi\bDelta\bPsi^\top$ and $\bPhi^\top\bPhi \approx \bI$, we have $$\bPsi \bDelta \bPsi^\top = \bPsi \bPhi^\top\bPhi \bDelta \bPsi^\top \approx \bPsi \bPhi^\top \bA \ . $$ Thus we may approximate the gradient  as
$$ \nabla L(\bx) \ \approx \ \bPsi  \bPhi^\top\Big( \bA\bx-\by \Big) $$
Also observe that $\bA\bPsi \bDelta^{\dagger} \approx \bPhi$, where $\bDelta^{\dagger}$ is the pseudoinverse of $\bDelta$ (i.e., $\bDelta^{\dagger}$ has diagonal elements $\delta_i^{-1}$ if $\delta_i>0$ and $0$ if $\delta_i=0$). Thus, we have $\bPsi  \bPhi^\top\approx \bPsi  \bDelta^{\dagger} \bPsi^\top \bA^\top$.   This shows that the gradient in (\ref{grad}) is approximately the gradient of $\tfrac{1}{2}\|\by-\bA\bx\|_2^2$ filtered by $\bPsi  \bDelta^{\dagger} \bPsi^\top$
\begin{eqnarray}
\nabla L(\bx) & \approx & \bPsi  \bDelta^{\dagger} \bPsi^\top \bA^\top\Big(\bA\bx- \by \Big) \ .
\label{filtered_grad}
\end{eqnarray}
The filtering operation $\bPsi  \bDelta^{\dagger} \bPsi^\top$ is a $\bPsi$-domain filtering operation that scales each component of the gradient according to the inverse attenuation factor of $\bA$ acting on the component.  Approximation (\ref{filtered_grad}) will be our preferred expression for the gradient.
   This filtering operation can be computed efficiently whenever $\bPsi$ admits a fast transform (e.g., Fourier or wavelet transform).  This leads to a novel iterative denoising algorithm (explained in Algorithm 2 below) called \emph{Filtered IDA}, where the \bb{\small denoise} step applies a soft-threshold to the coefficients $\bPsi^\top\widetilde \bx^{k+1}$ at threshold $\lambda$ and then computes the inverse transform by applying $\bPsi$ to the result.  Specifically, 
$$\bb{\small denoise}(\widetilde \bx^{k+1},\gamma) \ = \ \sum_{j} \st_{\gamma\lambda}\big(\bpsi_j^\top \widetilde\bx^{k+1}\big) \, \bpsi_j \ , $$
where $\st_{\gamma\lambda}$ is the soft-thresholding function (\ref{soft-threshold}) with threshold $\gamma\lambda$.
The key point is that this is the standard soft-threshold with a global threshold level, and there is no need to adjust the threshold for each coefficient separately to match the noise level induced by the linear operator.  In other words, this is an off-the-shelf soft-thresholding operation.  More generally, we may use any denoiser behaves similarly to the soft-thresholding operation.


\begin{figure}[h]
    \centering
    \includegraphics[width=8cm]{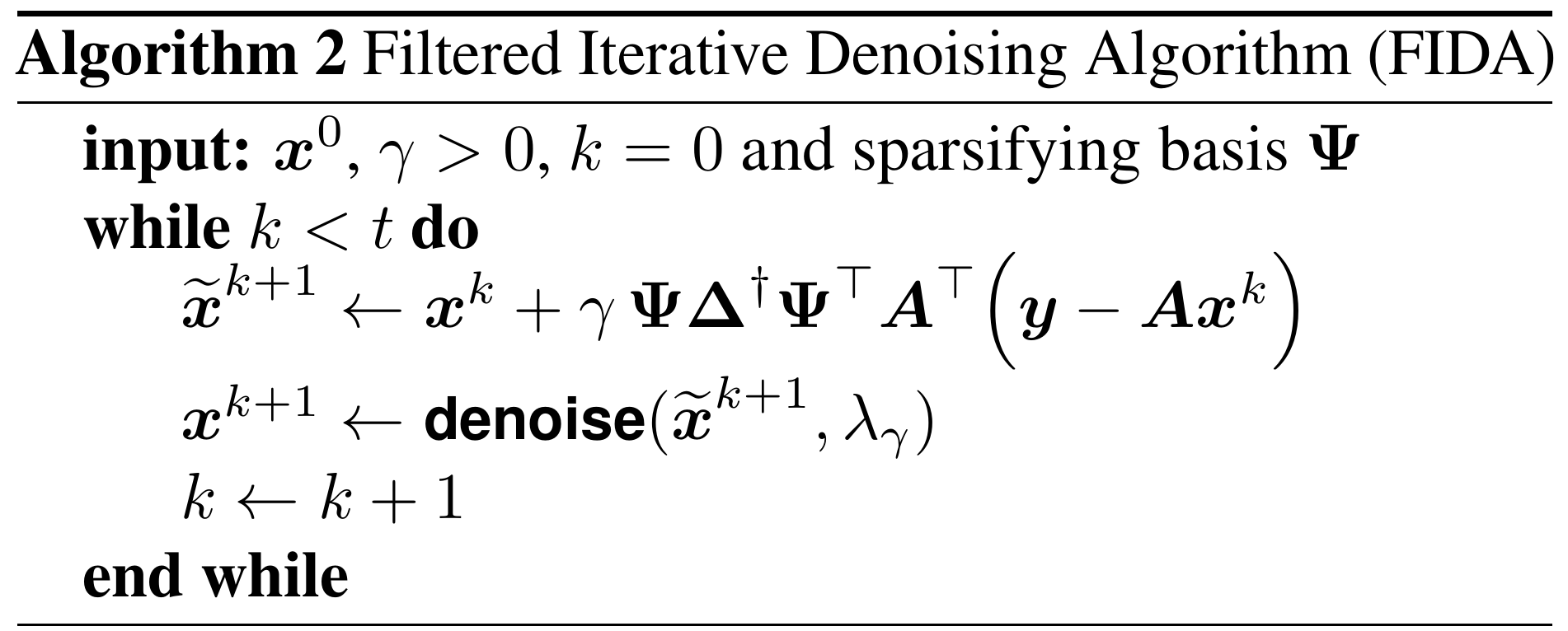} 
    \label{FIDA}
\end{figure}

\subsection{Choosing the Basis $\bPsi$}
The approach may be used in conjunction with an off-the-shelf denoising algorithm, represented by \bb{\small denoise} in FIDA.  The rationale of the approach rests on two considerations:
\begin{enumerate}
    \item The black-box denoiser operation is similar to soft-thresholding with an appropriate sparsifying basis  $\bPsi.$
    \item The operator $\bA$ satisfies Assumption~\ref{A1} with $\bPsi$.  
\end{enumerate}
The choice of $\bPsi$ may be based on either or both of these considerations. One can view $\bPsi$ as a hyperparameter of FIDA, and different options may be tried since the behavior of the black-box denoiser may be difficult to characterize.  We discuss several natural choices next.

{\bf Wavelet Basis:} Smooth wavelet bases are a good choice.  The theory of the \emph{wavelet-vaguelette decomposition} (WVD) \cite{donoho1995nonlinear,abramovich1998wavelet,chesneau2010stein} shows that if $\bPsi$ is an orthogonal wavelet transform and $\bA$ is a \emph{weakly invertible} linear operator like an integration operator or Radon transform operator, then Assumption~\ref{A1} is met.  Black-box denoisers trace their lineage back to wavelet and total variation denoising methods, so wavelet thresholding denoisers are a good candidate for approximating black-box denoisers as well.  The particular choice of wavelet maybe be viewed as a hyperparameter of the FIDA.  Smoother wavelets (i.e., not Haar wavelets) are suggested by the WVD theory and also recommended for the FIDA.

{\bf Learned Bases:} Bases or dictionaries learned from examples \cite{mairal2008learning} are also a potentially attractive choice. Learned representations may be non-orthogonal and redundant, but recall our main derivation hinges only on the $\bPhi$ (not $\bPsi$) being nearly orthogonal.  The diagonal elements of $\bDelta$ are determined by the norm of columns of $\bA\bPsi$, so in principle the filter $\bPsi \bDelta^{\dagger}\bPsi^\top$ may be effective even if $\bPsi$ is not orthogonal.

{\bf Diagonalizing Basis:} If $\bA = \bPsi\bDelta\bPsi^\top$ for some orthogonal matrix $\bPsi$, then the gradient in (\ref{filtered_grad}) is  $\bPsi\bD\bPsi^\top (\bA\bx-\by)$, where $\bD$ is a diagonal matrix with $1$ on each diagonal corresponding $\delta_i\neq 0$ and $0$ otherwise. This form is most easily obtained directly from (\ref{grad}).   Special instances of this setting occur when $\bA$ is diagonal and $\bPsi$ is the canonical basis or when $\bA$ is a (circular) convolution operator and $\bPsi$ is the Discrete Fourier Transform (DFT) basis. Of course, black-box denoisers may or may not be approximated by thresholding in these bases, but nonetheless may be attractive and perform well.

{\bf SVD Basis:} Let $\bA = \bPhi\bDelta\bPsi^\top$ denote the Singular Value Decomposition (SVD) of $\bA$. Then $\bA^\top \bPhi = \bPsi \bDelta$ and Assumption~\ref{A1} is satisfied (in fact $\bPhi$ is truly orthogonal in this case).  In this case, the the gradient in (\ref{grad}) is given by  $\bPsi\bPhi^\top(\bA\bx-\by)$. The SVD basis may or may not be good match for approximating the behavior of a given black-box denoiser.

\section{Experiments}
In our experiments, we use 10 grayscale images that are 256 x 256 in size with pixel values ranging $[0,255]$. We apply two different forward models ($\bA$ matrix) to our data. One is a Gaussian blur, and the other is variable sensor gain model, similar to what occurs in the satellite imaging problem described in Section~\ref{trouble}. We then add Gaussian noise with mean zero and standard deviation $\sigma$ to get our degraded images.

Because deblurring a noisy image is an ill-posed inverse problem, we assume a low-noise regime. We apply Gaussian noise with $\sigma = 0.2, 1, 5$ to our dataset. On the other hand, gain correction of a noisy image is well-posed, so we looked at a higher noise setting where $\sigma = 5, 10, 20$.

Using BM3D as our denoiser, we compare our algorithm against the standard IDA. We implement our method with the wavelet basis (W-FIDA), using the Daubechies D6 wavelet basis, as well as with the diagonalizing basis (D-FIDA). For the deblurring problem, the Fourier basis is the diagonalizing basis, and for gain correction, the diagonalizing basis is the pixel basis. 

\begin{figure*}
    \centering
    \includegraphics[width=5cm]{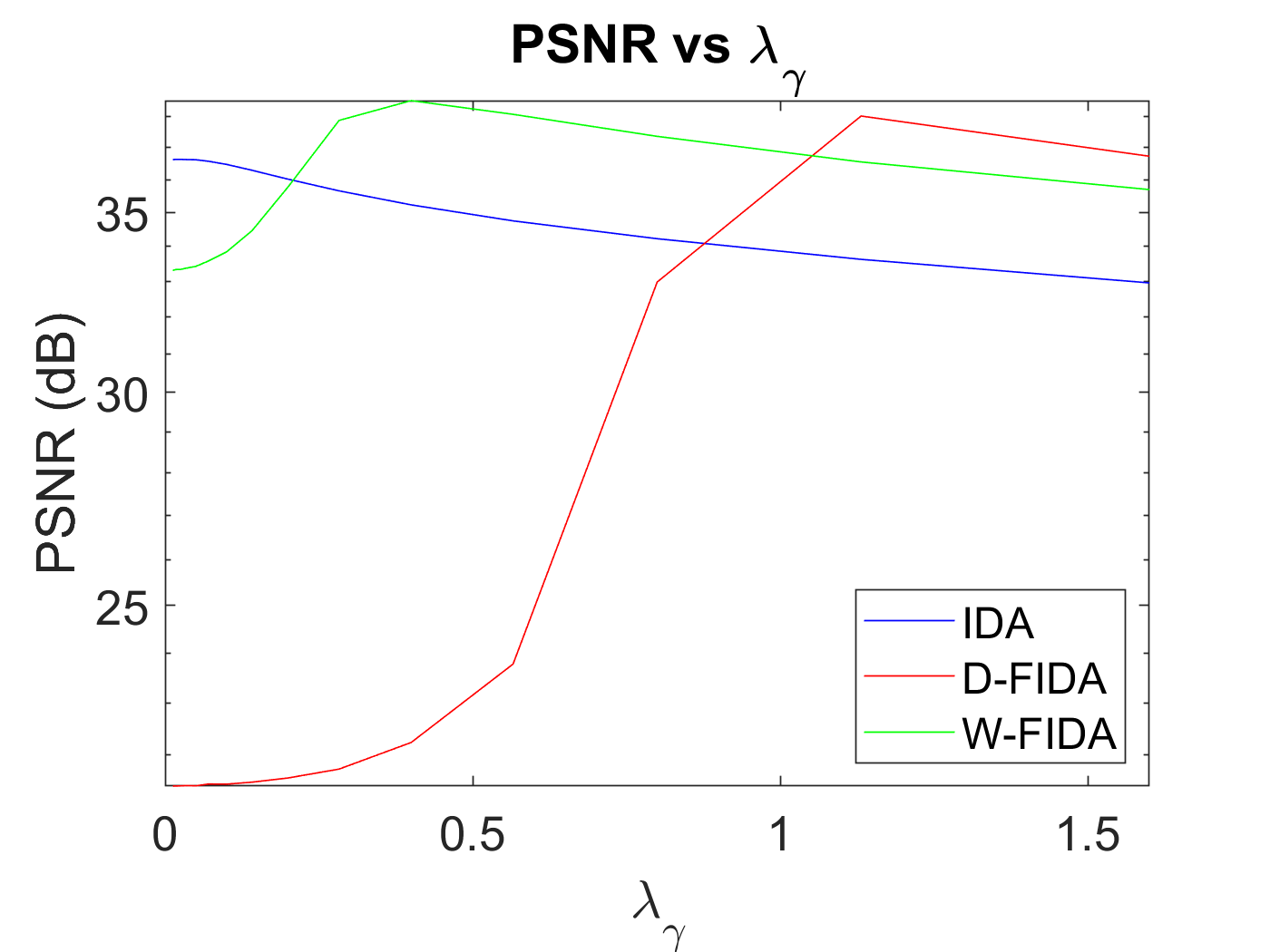}
    \includegraphics[width=5cm]{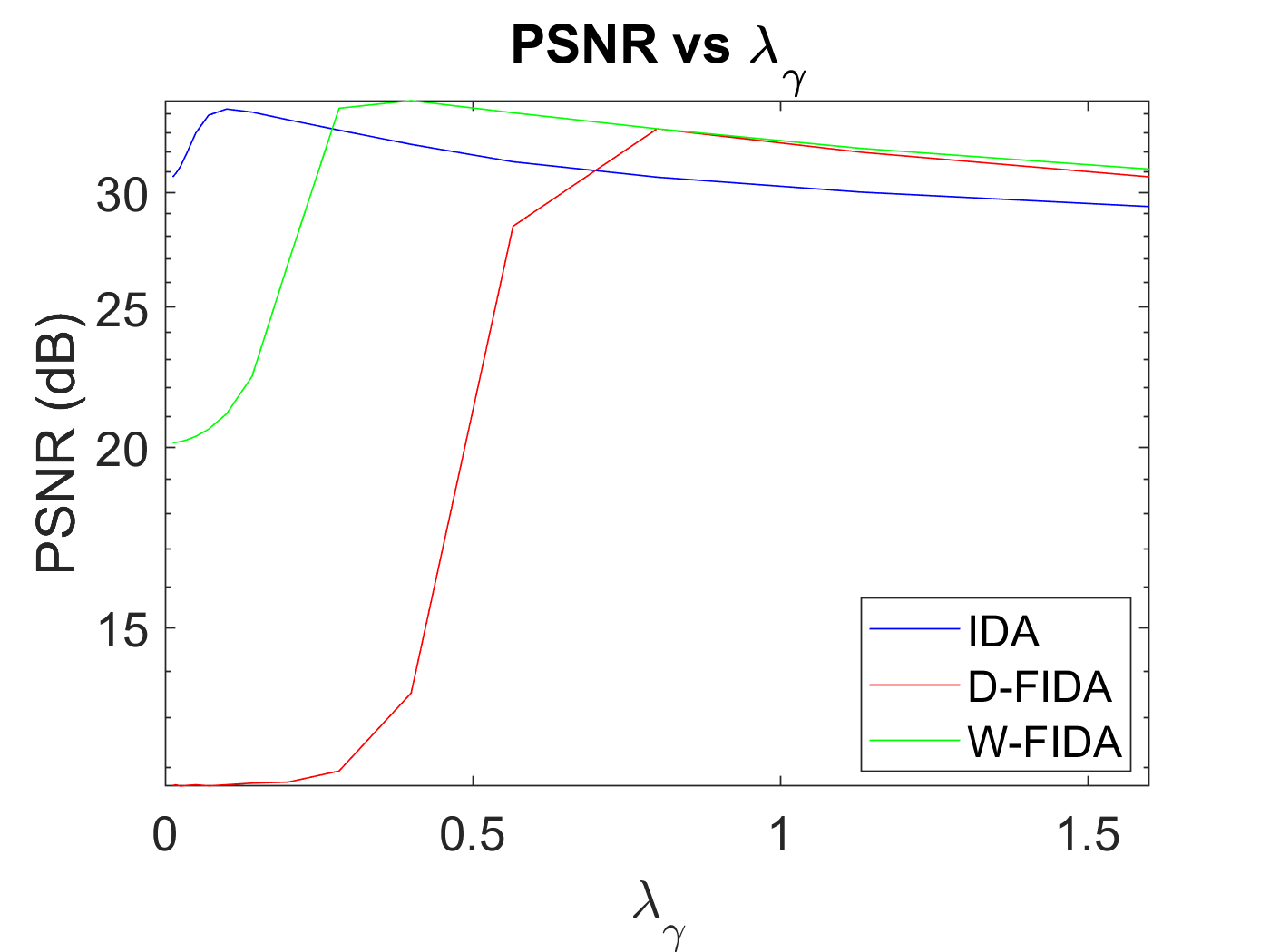}
    \includegraphics[width=5cm]{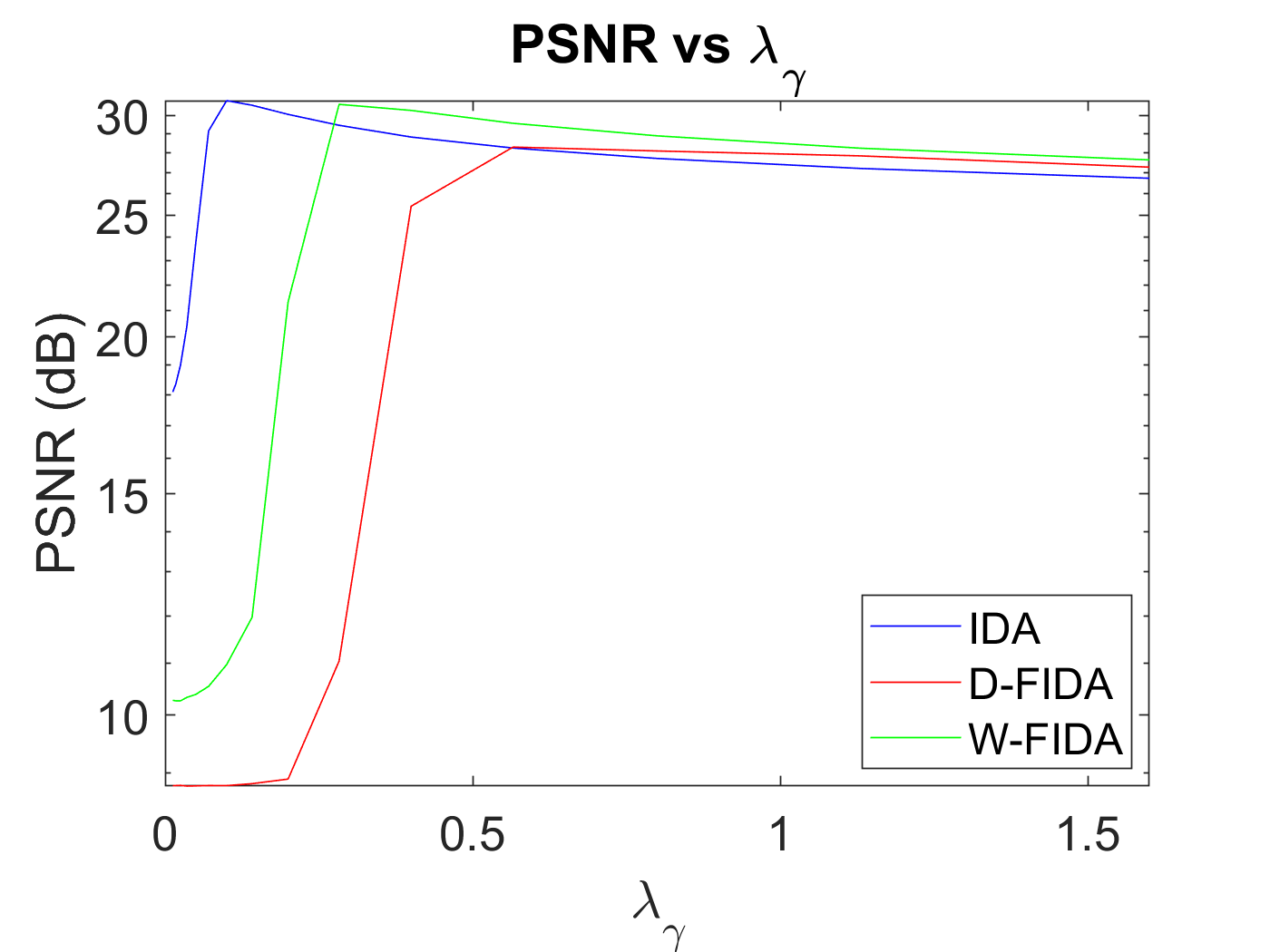}
    \caption{PSNR (dB) vs denoising parameter for blurred "hill" image. From left to right, the plots show results when $\sigma = 0.2$, $\sigma = 1$, and $\sigma = 5$. The results in Table~\ref{blur_bm3d} report the PSNR using the best value of $\lambda_\gamma$ for each method.}
    \label{fig:psnr_lambda}
\end{figure*}

For fair comparison, we average the peak-signal-to-noise-ratio (PSNR) measured in dB over 10 independent runs, and we sweep over a range of denoising parameters, $\lambda_\gamma$. The results in Tables \ref{blur_bm3d}-\ref{gain_bm3d} show the best average PSNR over $\lambda_\gamma$ for each method, as depicted in Figure~\ref{fig:psnr_lambda}.

\begin{figure}[h]
    \centering
    \includegraphics[width=4.4cm]{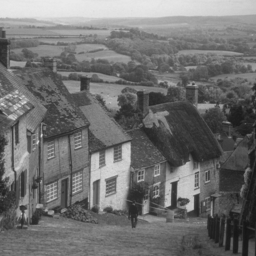} \\
    \includegraphics[width=1.4cm]{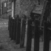}
    \includegraphics[width=1.4cm]{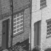}
    \includegraphics[width=1.4cm]{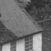}
    \caption{Image of hill groundtruth.}
\label{fig:hill}
\end{figure}

\begin{figure*}
    \centering
    \includegraphics[width=5cm]{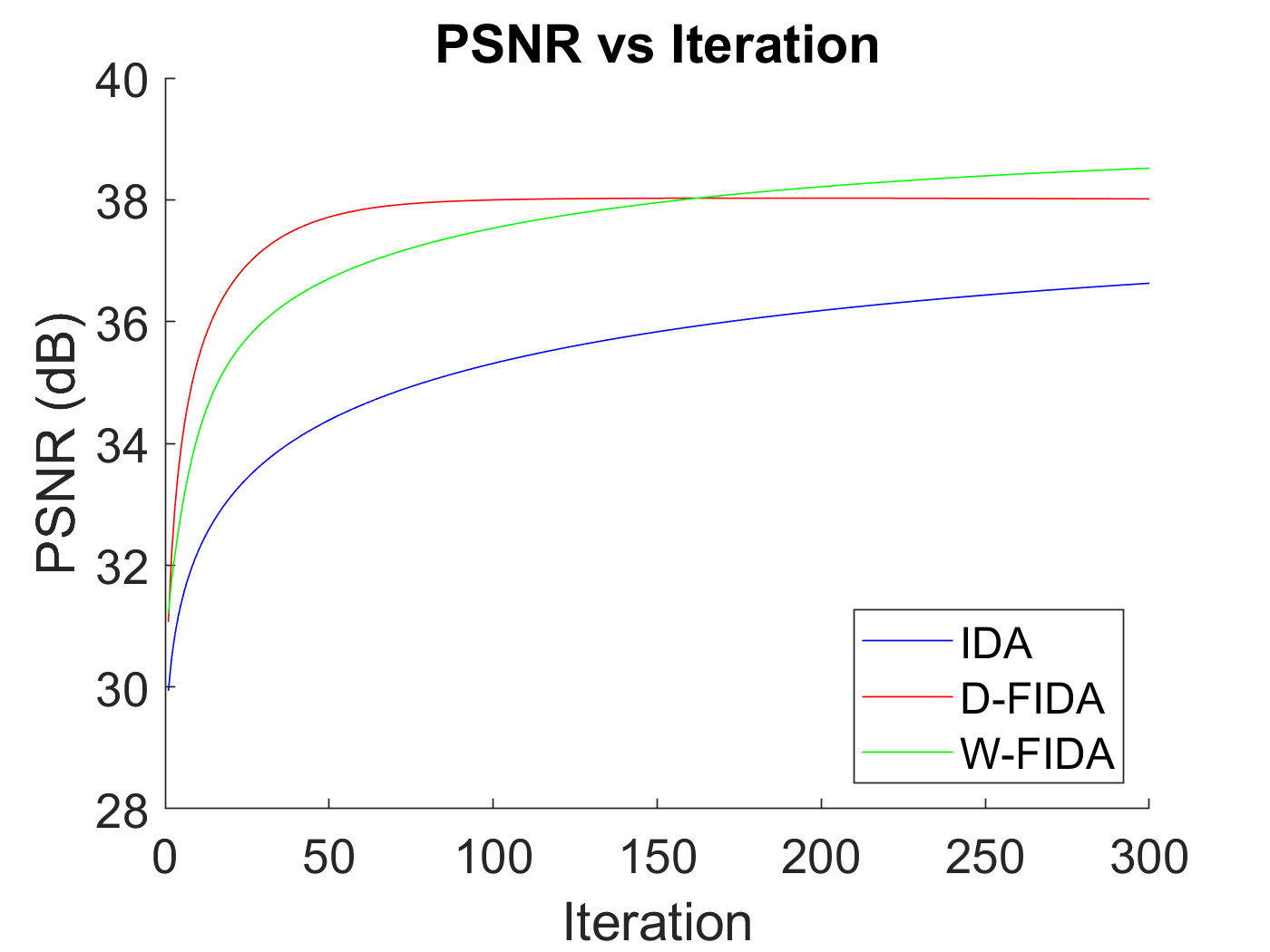}
    \includegraphics[width=5cm]{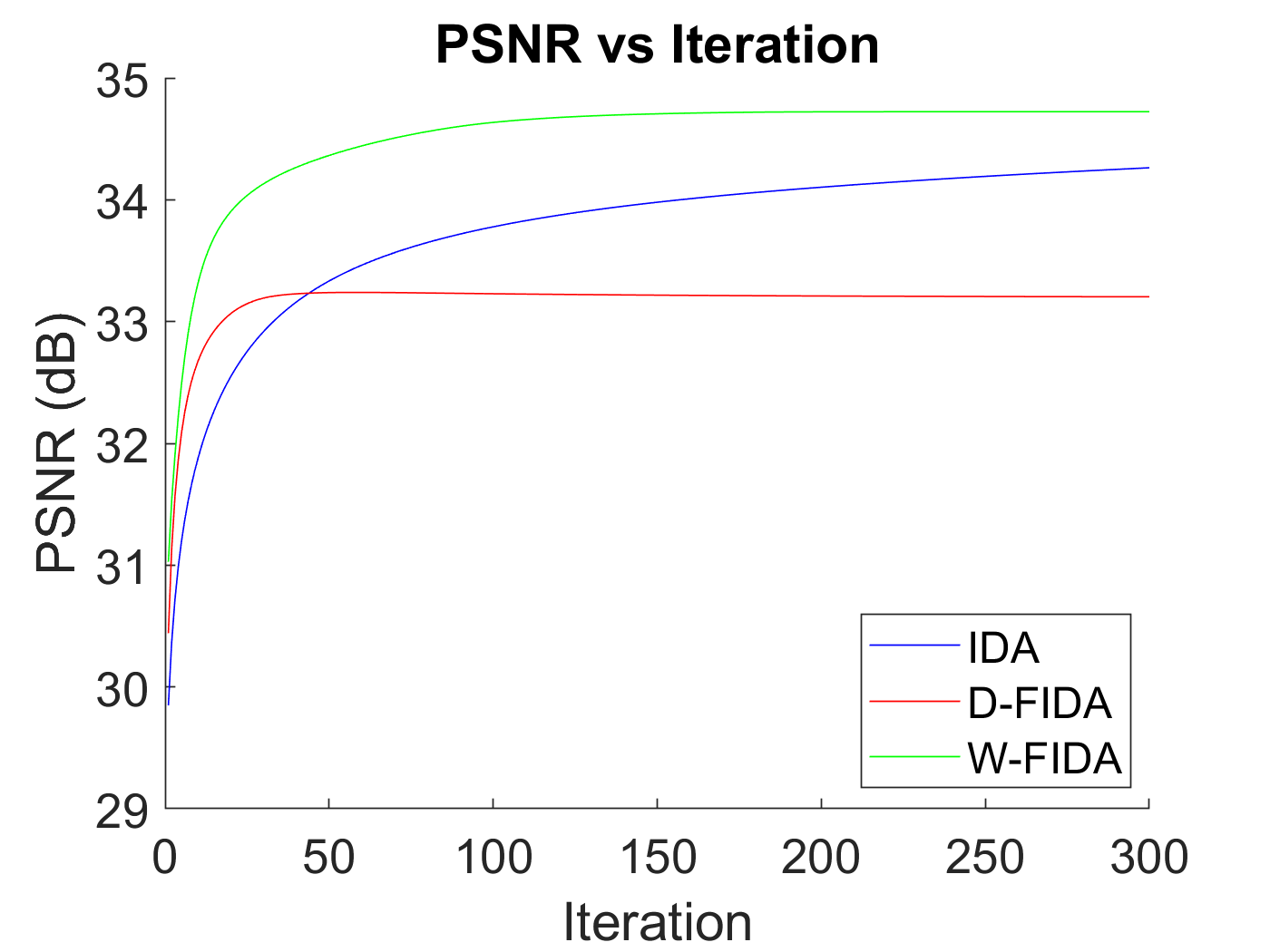}
    \includegraphics[width=5cm]{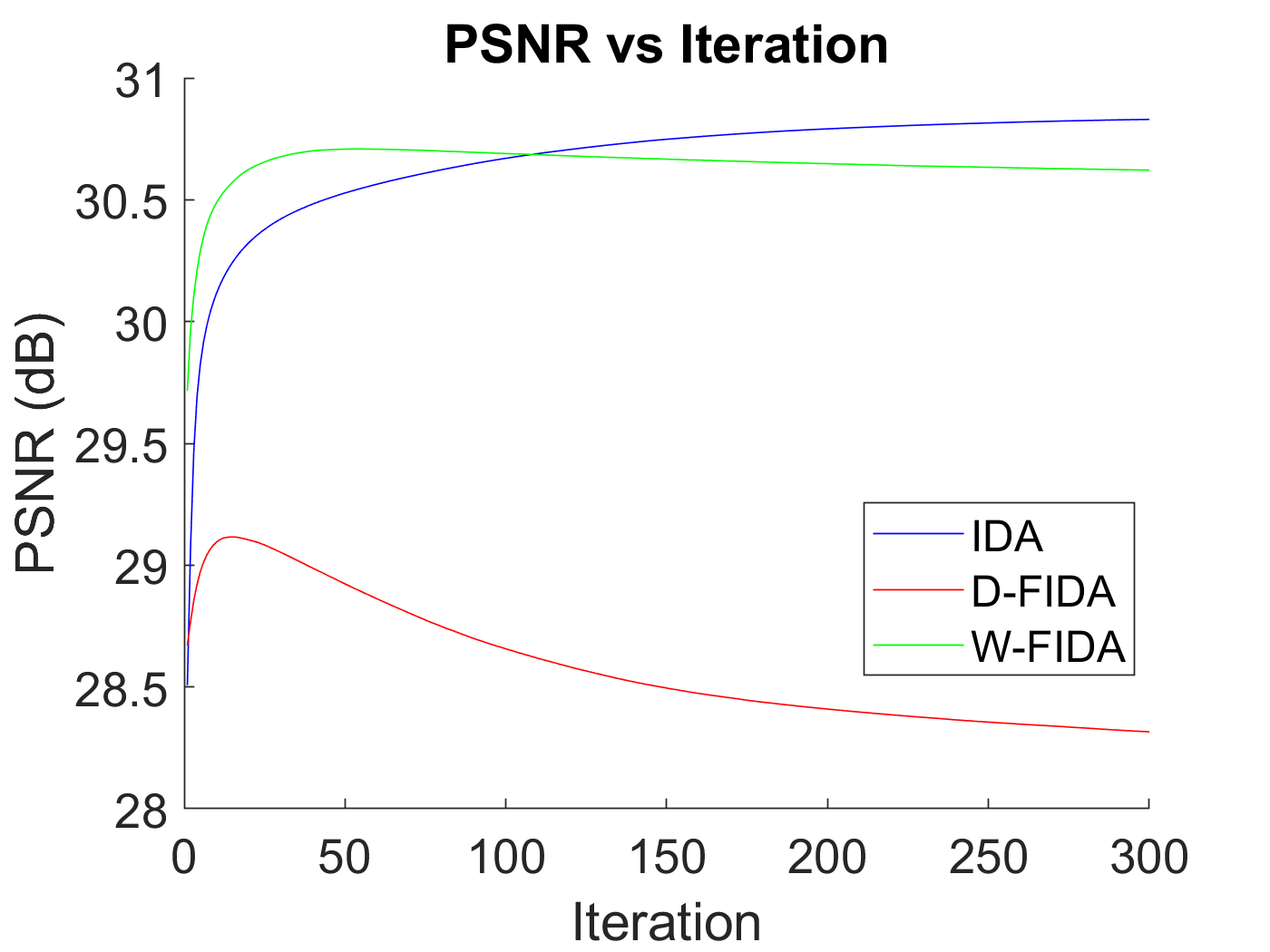}
    \caption{PSNR (dB) vs iteration for blurred "hill" image. From left to right, the plots show the PSNR per iteration when $\sigma = 0.2$, $\sigma = 1$, and $\sigma = 5$. The denoising parameter, $\lambda_\gamma$, for each method is assumed to be the $\lambda_\gamma$ that produced the best results.}
    \label{fig:psnr_iter}
\end{figure*}

\begin{figure*}
    \centering
    \includegraphics[width = 17cm]{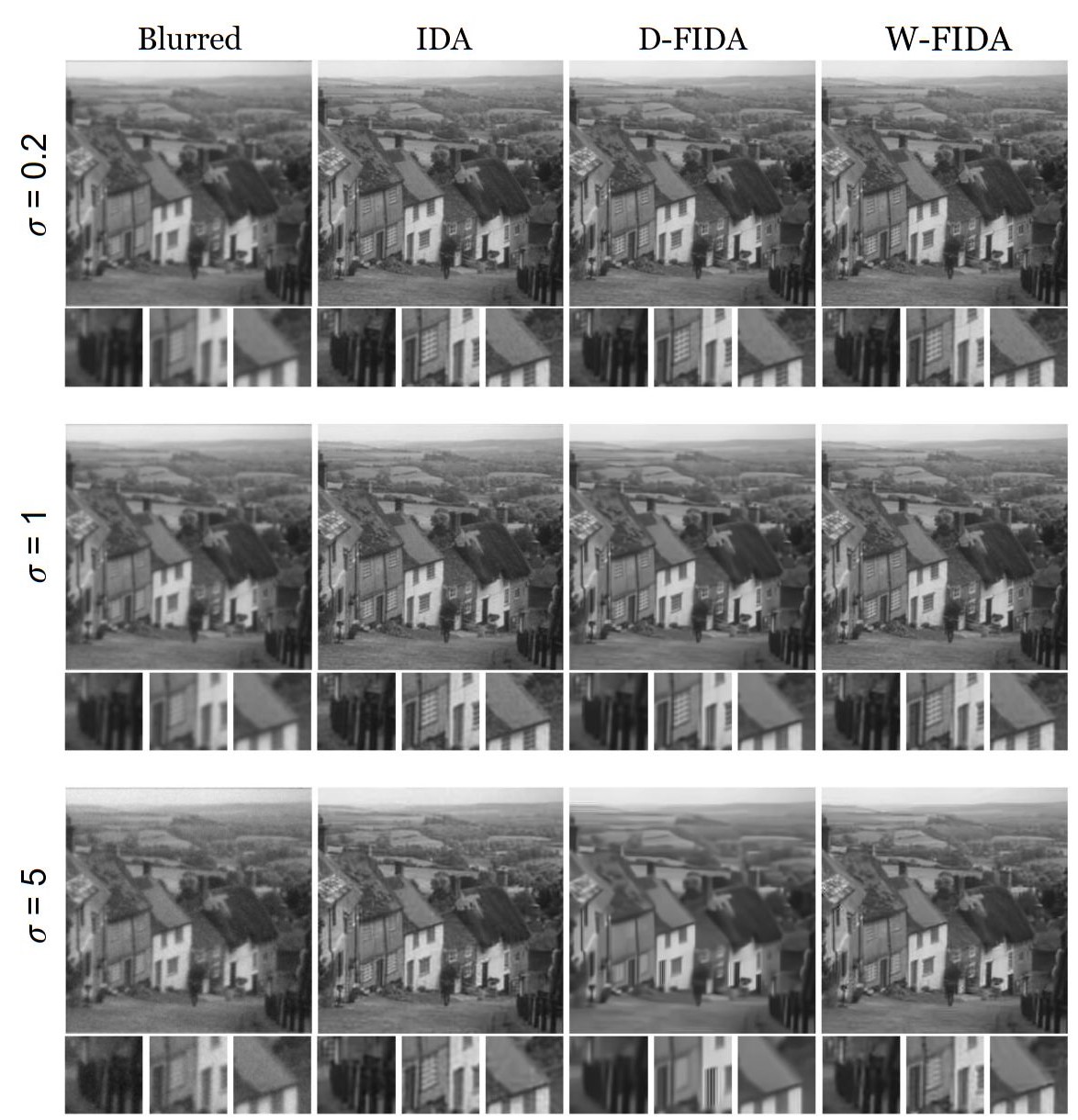}
    \caption{Noisy deblurred images. From left to right we have the blurry and noisy image, denoised by IDA, denoised by D-FIDA, and denoised by W-FIDA. From top to bottom we have $\sigma = 0.2$, $\sigma = 1$, and $\sigma = 5$.}
    \label{fig:denoised_image_block}
\end{figure*}

In Figures~\ref{fig:psnr_lambda}-\ref{fig:denoised_image_block}, we look more closely at a particular example, the blurred "hill" image; the original is shown in Figure~\ref{fig:hill}. 


\begin{table*}[]
    \caption{PSNR (dB) for Deblurring Problem}
    \label{blur_bm3d}
    \centering
    \small{
    \begin{tabular}{|c|c|c|c|c|c|c|c|c|c|c|}
        \hline
        & boats & bridge & camera & couple & flag & hill & house & man & peppers & saturn \\
        \hline
        \multicolumn{11}{|c|}{$\sigma$ = 0.2} \\
        \hline
         IDA & 36.2593   &31.3291	&32.7814	&36.1147	&32.8976	&36.6289	&38.9325	&35.8846	&31.7055	&48.9936  \\
         \hline
         D-FIDA & 38.283	&32.8457	&\textbf{36.1399}	&38.2633	&\textbf{37.0899}	&38.0141	&40.7933	&36.9305	&\textbf{37.7321}	&50.6108 \\
         \hline
         W-FIDA & \textbf{38.4229}	&\textbf{32.9955}	&35.4155	&\textbf{38.4624}	&35.9628	&\textbf{38.5186}	&\textbf{41.6699}	&\textbf{37.4383}	&35.982	&\textbf{51.5752} \\
         \hline
         \multicolumn{11}{|c|}{$\sigma$ = 1} \\
         \hline
         IDA & 33.7121	&29.5842	&31.5512	&33.7485	&32.1291	&34.2627	&36.5864	&33.363	&31.4105	&47.5211\\
         \hline
         D-FIDA & 32.8805	&28.8595	&31.6366	&32.9861	&33.3814	&33.2032	&35.9584	&32.2233	&34.3249	&46.9006\\
         \hline
         W-FIDA &\textbf{34.0971}	&\textbf{29.7239}	&\textbf{32.3855}	&\textbf{34.0051}	&\textbf{33.6895}	&\textbf{34.7233}	&\textbf{37.0234}	&\textbf{33.5791}	&\textbf{34.5936}	&\textbf{47.8242}\\
         \hline
         \multicolumn{11}{|c|}{$\sigma$ = 5} \\
         \hline
         IDA & \textbf{29.9958}	&\textbf{26.869}	&\textbf{29.0645}	&\textbf{30.0964}	&30.0994	&\textbf{30.8298}	&\textbf{34.0729}	&\textbf{29.776}	&30.6765	&42.647\\
         \hline
         D-FIDA & 28.0383	&25.2469	&27.7563	&27.5693	&29.0831	&28.3158	&32.5683	&28.0429	&30.231	&42.0463\\
         \hline
         W-FIDA & 29.8612	&26.7151	&29.0512	&29.8505	&\textbf{30.1599}	&30.6215	&34.0443	&29.6727	&\textbf{31.4304}	&\textbf{42.9068} \\
         \hline
    \end{tabular}
    }
\end{table*}

\begin{table*}[]
    \caption{PSNR (dB) for Gain Correction Problem}
    \label{gain_bm3d}
    \centering
    \small{
    \begin{tabular}{|c|c|c|c|c|c|c|c|c|c|c|}
        \hline
        Image & boats & bridge & camera & couple & flag & hill & house & man & peppers & saturn \\
        \hline
        \multicolumn{11}{|c|}{$\sigma$ = 5} \\
        \hline
         IDA & 35.8209	&\textbf{33.8451}	&37.0098	&36.218	&37.401	&\textbf{34.7144}	&\textbf{38.4773}	&\textbf{34.5367}	&\textbf{36.9739}	&\textbf{45.1132}\\
         \hline
         D-FIDA & \textbf{36.0732}	&33.736	&\textbf{37.2867}	&36.2702	&\textbf{37.6331}	&34.4821	&38.3447	&34.2866	&36.6743	&44.9589 \\
         \hline
         W-FIDA & 36.0264	&33.6706	&37.2478	&\textbf{36.281}	&37.6173	&34.5156	&38.3617	&34.3021	&36.6795	&44.8634\\
         \hline
         \multicolumn{11}{|c|}{$\sigma$ = 10} \\
         \hline
         IDA & 31.8819	&29.2484	&33.1915	&31.9839	&34.7869	&30.7138	&35.1479	&30.1071	&33.1833	&\textbf{41.91}\\
         \hline
         D-FIDA & \textbf{32.1274}	&\textbf{29.55}	&\textbf{33.4604}	&\textbf{32.4706}	&\textbf{34.8465}	&\textbf{31.2209}	&\textbf{35.1966}	&\textbf{30.4129}	&33.4513	&41.8671\\
         \hline
         W-FIDA &32.108	&29.5456	&33.4544	&32.4692	&34.7861	&31.1827	&35.1856	&30.4106	&\textbf{33.4597}	&41.715\\
         \hline
         \multicolumn{11}{|c|}{$\sigma$ = 20} \\
         \hline
         IDA & 28.4247	&25.9532	&\textbf{29.7566}	&28.5431	&31.4189	&27.7433	&\textbf{32.2804}	&26.5139	&29.7204	&38.1508\\
         \hline
         D-FIDA & \textbf{28.5858}	&\textbf{26.0796}	&29.6381	&\textbf{28.9446}	&\textbf{31.7539}	&\textbf{28.0874}	&32.2439	&\textbf{27.1913}	&\textbf{30.1773}	&\textbf{38.3859}\\
         \hline
         W-FIDA & 28.5386	&26.0367	&29.6253	&28.904 	&31.6136	&28.0541	&32.2276	&27.186 	&30.1156	&38.2836\\
         \hline
    \end{tabular}
    }
\end{table*}


\section{Discussion}
From the results in Table~\ref{blur_bm3d}, we see that our proposed filtered IDA method used for deblurring significantly outperforms the standard IDA when noise is low. When the standard deviation of the noise is high ($\sigma = 5$), the standard IDA slightly outperforms our method in some cases, in terms of PSNR. However, visual inspection of the example in Figure~\ref{fig:denoised_image_block}  suggests that the proposed filtered IDA removes noise better than standard IDA while still maintaining qualitatively good structure. Standard IDA appears to opt towards keeping noise, avoiding error due to inaccurate reconstruction. This may explain why the standard IDA has a slightly higher PSNR for the largest level of $\sigma$.

Table~\ref{blur_bm3d} also shows W-FIDA consistently outperforming D-FIDA in the deblurring problem. By looking at Figure~\ref{fig:psnr_iter}, which shows the PSNR as a function of IDA iterations, we can speculate why. When using the diagonalizing basis, the PSNR peaks after a small number of iterations and then worsens with further iterations.  The peak PSNR is also lower than that of the other IDA methods.  Behavior like this suggests that the performance of IDA/FIDA may be enhanced by additional regularization in form of early stopping.

When considering the gain correction problem, we see a pattern in Table~\ref{gain_bm3d} opposite to that in the deblurring setting. That is, as $\sigma$ increases, our proposed method begins to outperform standard IDA. Furthermore, our diagonalizing basis is consistently outperforming our wavelet basis, though the two bases do tend towards the same PSNR in most cases.

\begin{figure*}
    \centering
    \includegraphics[width=15cm]{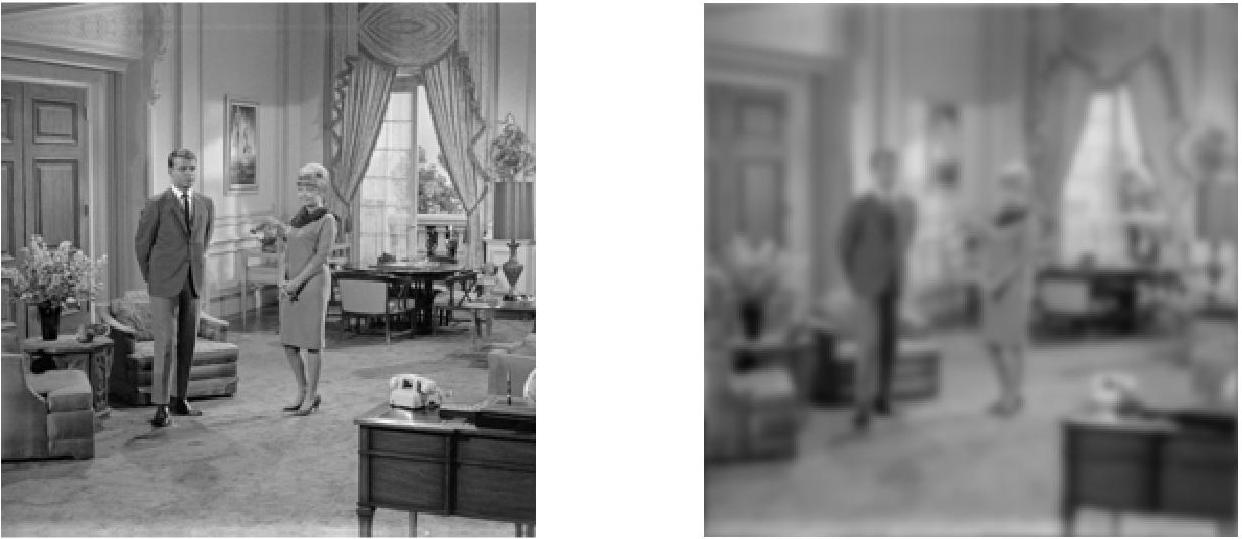} \\ \ \\
    \includegraphics[width=15cm]{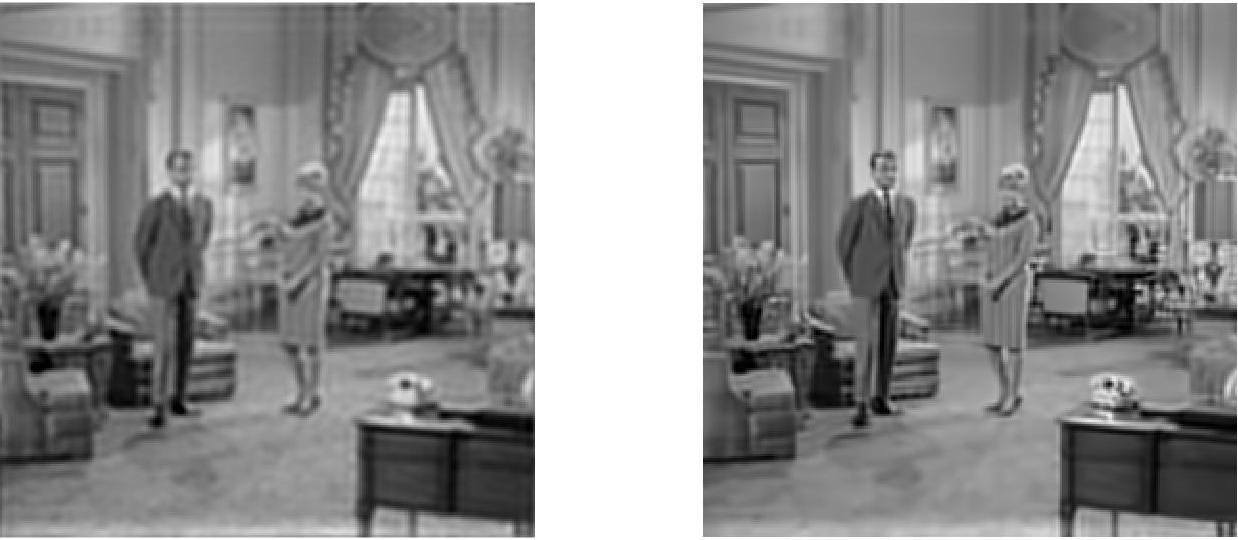}
    \caption{Severe deblurring example.  Original and blurred image (top), PSNR 26.5dB.  Standard IDA (bottom right) and proposed Filtered IDA (bottom left), PSNR = 28dB. The regularization parameter was adjusted to obtain the highest PSNR in each case.}
    \label{fig:couple}
\end{figure*}

The visual distinctions in Figure~\ref{fig:denoised_image_block} are small due to the high PSNR.  More striking visual differences occur in more severe blurring situations, like that illustrated in Figure~\ref{fig:couple}.  In this experiment, we use a strong Gaussian blur with a small amount of noise and compare standard IDA to FIDA, both using BM3D.  FIDA uses the diagonal (Fourier basis) filtering, in this example.  The regularization parameter was adjusted to produce the best result for each method (for fair comparison of the best performances).  FIDA provides a visually better result with significantly higher PSNR.  Interestingly, both IDA and FIDA introduce a slight striping artifact in the woman's dress, which we believe is related to the BM3D denoiser.

\section{Conclusions}

This paper proposes a simple linear filtering operation for gradient updates in iterative denoising algorithms for linear inverse problems.  The filtering accounts for the specific linear operator involved in the problem, and eliminates the need to adapt the denoiser to the problem.  The derivation of the Filtered Iterative Denoising Algorithm (FIDA) is based on a denoising by thresholding in an appropriate transform domain, such as the wavelet domain.  This may be a reasonable approximation to the function of many black-box denoisers, which motivates the use of FIDA with a variety of denoising methods.

There are a number of interesting directions for future research.  First and foremost is developing a deeper understanding of how the choice of the basis $\bPsi$ used for filtering interacts with specific black-box denoisers and how this effects overall performance.  Recall that the filtering operation multiplies the usual gradient $\bA^\top (\bA\bx-\by)$ by the matrix $\bPsi \bDelta^{\dagger}\bPsi$.  More generally, arbitrary matrices could be used to filter the gradient.  One generalization of the current filter $\bPsi \bDelta^{\dagger} \bPsi^\top$ is to replace $\bDelta^{\dagger}$ with another diagonal matrix.  Recall that the diagonal elements are $\delta_i^{-1}$, if $\delta_i\neq 0$, and $0$ otherwise.  Instead we could set the diagonal elements to be $\delta_i / (|\delta_i|^2 + \tau)$ with a small $\tau>0$, in the spirit of Wiener filtering. The parameter $\tau$ plays the role of an additional hyperparameter that can be tuned to improve performance. The filtering matrix could even be learned from data for a particular inverse problem domain, which is arguably simpler that adapting or learning a black-box denoiser from scratch.  Finally, our experiments focused on problems in satellite image denoising in the presence of heterogeneous sensor gains and image deblurring/deconvolution.  Future work should explore the potential of FIDA in a broader range of applications, possibly including tomographic reconstruction and superresolution.  This paper also focuses on the squared error loss and a particular form of iterative denoising. The filtering approach proposed here might be applied to other formulations such as regularization-by-denoising (RED) \cite{romano2017little}, deep unfolding \cite{chen2018theoretical}, and multiagent consensus equilibrium (MACE) \cite{buzzard2018plug,pmlr-v97-ryu19a,hurault2022gradient}.




\newpage

\bibliography{refs}
\bibliographystyle{icml2022}



\end{document}